\documentclass[manuscript]{aastex}

\usepackage{bm}
\usepackage{multirow}
\usepackage{color}
\usepackage{ulem}

\newcommand{\beq}{\begin{equation}}
\newcommand{\eeq}{\end{equation}}
\newcommand{\beqa}{\begin{eqnarray}}
\newcommand{\eeqa}{\end{eqnarray}}

\shorttitle{Acoustic cutoff period in the solar atmosphere}
\shortauthors{Murawski {\et al.}}

\begin{document}


\title{Variation of Acoustic Cutoff Period with Height in the Solar 
Atmosphere: Theory versus Observations}


\author{K. Murawski$^{1}$, Z. E. Musielak$^{2,3}$, P. Konkol$^{1}$, and A. Wi\'sniewska$^{3}$}

\affil{$^{1}$University of Maria Curie-Sk{\l}odowska, Institute of Physics, Group of Astrophysics, 
Radziszewskiego 10, PL -- 20 031 Lublin, Poland;} 
\affil{$^2$Department of Physics, University of Texas at Arlington, Arlington, TX 76019, USA;}
\affil{$^{3}$Kiepenheuer-Institut f\"ur Sonnenphysik, Sch\"oneckstr. 6, 79104 Freiburg, Germany}
\email{kmur@kft.umcs.lublin.pl; zmusielak@uta.edu; piotrk@kft.umcs.lublin.pl; 
wisniewska@leibniz-kis.de}

\begin{abstract}
Recently Wi\'sniewska et al. (2016) demonstrated observationally how the acoustic cutoff frequency
varies with height in the solar atmosphere that included the upper photosphere and the lower and 
middle chromosphere, and showed that the observational results cannot be accounted for by the 
existing theoretical formulas for the acoustic cutoff.  In order to reproduce the observed 
variation of the cutoff with the atmospheric height, numerical simulations of impulsively generated 
acoustic waves in the solar atmosphere are performed, and the spectral analysis of temporal wave 
profiles is used to compute numerically changes of the acoustic cutoff with height. Comparison 
of the numerical results to the observational data shows a good agreement, which clearly indicates 
that the obtained results may be used to determine the structure of the background solar atmosphere.  
\end{abstract}



\section{Introduction}
%
Propagation of acoustic waves in the solar atmosphere has been a subject of 
many analytical and numerical studies over several decades.  The main goal 
of these studies has been to understand the wave energy transfer from the 
solar convection zone, where the waves are generated, to the solar atmosphere,
where they may dissipate their energy and heat the background atmosphere.  A 
concept of acoustic cutoff frequency has played an important role in these 
studies, as it is this cutoff frequency that uniquely determines the propagation
conditions for acoustic waves in the solar atmosphere. 

The acoustic cutoff (period) frequency was originally introduced by Lamb (1909, 1910), 
who first considered an isothermal atmosphere and showed that the resulting cutoff
frequency is global (the same in the entire atmosphere) and is defined as the ratio 
of sound speed to twice the pressure or density scale heights, which are the same. 
Lamb (1910, 1932) extended his studies of acoustic waves to an atmosphere with uniform
temperature gradients and demonstrated how to define the acoustic cutoff in such a 
non-uniform medium.  Lamb's work was followed by many, however, specific solar physics
related applications were done by Moore \& Spiegel (1964), Souffrin (1966), Summers
(1976), Campos (1986), Gough (1993), Fleck \& Schmitz (1993), and more recently by
Musielak et al. (2006), Fawzy \& Musielak (2012) and Routh \& Musielak (2014).  
Different expressions for the cutoff were derived analytically and used in different 
studies of acoustic waves in the solar atmosphere.  However, a recent work by Wi\'sniewska 
et al. (2016) clearly demonstrated that these analytically obtained formulas did fail 
to account properly for the observed variation of the cutoff with height in the solar 
atmosphere reported by these authors.  

There are also a number of numerical studies of the acoustic wave propagation in the
solar atmosphere (e.g., Ulmschneider 1971; Ulmschneider et al. 1978; Carlsson et al. 
1997; Cuntz et al. 1998; Fazwy et al. 2002) in which the authors tried to determine 
the role played by acoustic waves of different frequencies in the atmospheric heating.  
Typically propagating acoustic waves were considered, which means that the authors did 
not have to dwell upon the concept of the acoustic cutoff frequency.  Variations of the 
acoustic cutoff frequency with height in the solar atmosphere were just recently 
reported by Wi\'sniewska et al. (2016), who performed observations using the 
Helioseismological Large Regions Interferometric Device operating at the Vacuum Tower 
Telescope located on Tenerife. The paper shows clear observational evidence for the 
existence of the cutoff in the solar atmosphere and its variation with the atmospheric
height.  In previous work, Jim\'enez (2006) and Jim\'enez et al. (2011) presented 
variations of the cutoff with a solar cycle.  

The main goal of this paper is to perform numerical simulations of impulsively generated 
acoustic waves in the solar atmosphere, and use the spectral analysis of temporal wave 
profiles to calculate numerically variations of the acoustic cutoff frequency with height.  
The obtained numerical results are compared to the observational data and with a good 
agreement between the theory and data, it is concluded that the results of this paper 
may become a basis for using the waves to determine the structure of the background 
solar atmosphere. 

This paper is organized as follows: our model of the solar atmosphere and numerical 
results are presented in Sects.~2 and 3, respectively; our conclusions are given in
Sect.~4.   
%
\section{Model of the solar atmosphere}\label{sec:numerical_model}
Our one-dimensional model of the solar atmosphere contains a gravitationally-stratified 
and magnetic field-free plasma, which is described by the Euler equations with the adiabatic 
index $\gamma=1.4$, the gravity  $g=(0,-g,0)$ and its solar value $g=274$ m s$^{-2}$, 
and a mean particle mass $m$ specified by a mean molecular weight $1.24$. 
Our assumption of one-dimensionality can be justified as we consider acoustic waves propagation over the 
atmospheric height of $1$ Mm.  This height is comparable with average size of a solar 
granule which is taken as a source of the waves.  

As we aim to study a quiet solar region, we assume that initially, at time $t=0$ s, low layers of the solar atmosphere are magnetic-free and 
they are in static equilibrium (with velocity $ V=0$) 
in which the equilibrium mass density and gas pressure 
are specified by a realistic, semi-empirical model of the plasma temperature $T(y)$ 
developed by Avrett \& Loeser (2008). 

The atmospheric equilibrium described above is perturbed 
by a Gaussian 
pulse in the vertical, $y$-component of the velocity given by
\beq
\label{eq:init_per}
V_{\rm y}(y,t=0) = A_{\rm v} \exp\left[ 
-\left(\frac{y-y_{\rm 0}}{w_{\rm y}}\right)^2  
\right]\, ,
\eeq
where $y$ is the vertical coordinate, 
$A_{\rm v}$ is the amplitude of the pulse, $y_{\rm 0}$ is its initial 
position, $w_{\rm y}=50$ km denotes its width along the vertical direction. 
This initial pulse corresponds to a packet of waves of its Gaussian spectrum 
characterized a wavenumber $k$. Since locally different $k$ corresponds to a 
different cyclic frequency $\omega$, we actually have present a packet of waves with different $\omega$. 
Once this packet propagates through the solar atmosphere, the atmosphere filters those wave frequencies 
that correspond to propagating acoustic waves; waves that become evanescent do not show up at higher atmospheric heights. 
It is this very characteristic behavior of the waves that is considered 
here to determine variations of the acoustic cutoff with height, 
and compare the numerically obtained wave periods to the observational data reported by Wi\'sniewska et al. (2016). 
%
\section{Numerical results}\label{sec:num_sim_model}
%
We solve numerically equations of hydrodynamics by using the PLUTO code 
in which we adopted HLLD Riemann solver and minmod flux-limiter (Mignone et al. 2012). 
Numerical simulations are performed 
in the model of the solar atmosphere described in Sect.~2. 
The simulation region is set as $-0.5 < y < 40$ Mm. 
At the bottom and top boundaries we set all plasma quantities to their equilibrium values. 
The region $-0.5 < y < 6.68$ Mm is covered by $1536$ uniform grid points, 
while the top level is represented by $512$ growing in size with height numerical cells. 
Such a stretched grid works as a sponge absorbing the incoming signal, 
and it results in negligibly small wave reflection from the top boundary. 
This very long domain, and the boundary type 

are not relevant for this simulation, 
because 
although the waves reach the upper boundary within the time range of interest 
they are strongly diffused in the top region, 
and therefore they do not affect the wave behaviour below the transition region. 

\begin{figure*}[h!]
\begin{center}
\vspace*{-2cm}
\includegraphics[width=8.175cm]{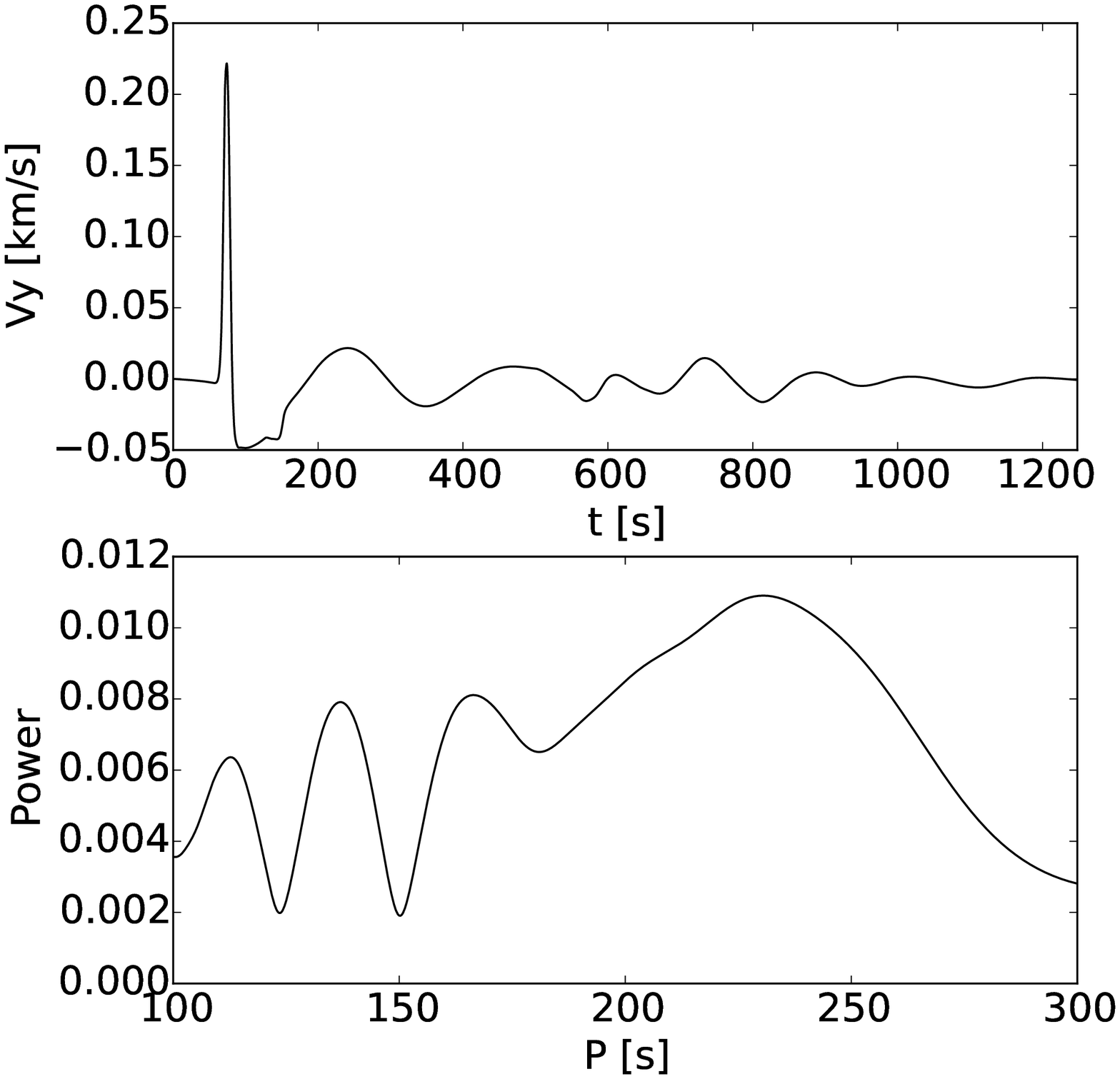}
\includegraphics[width=8.175cm]{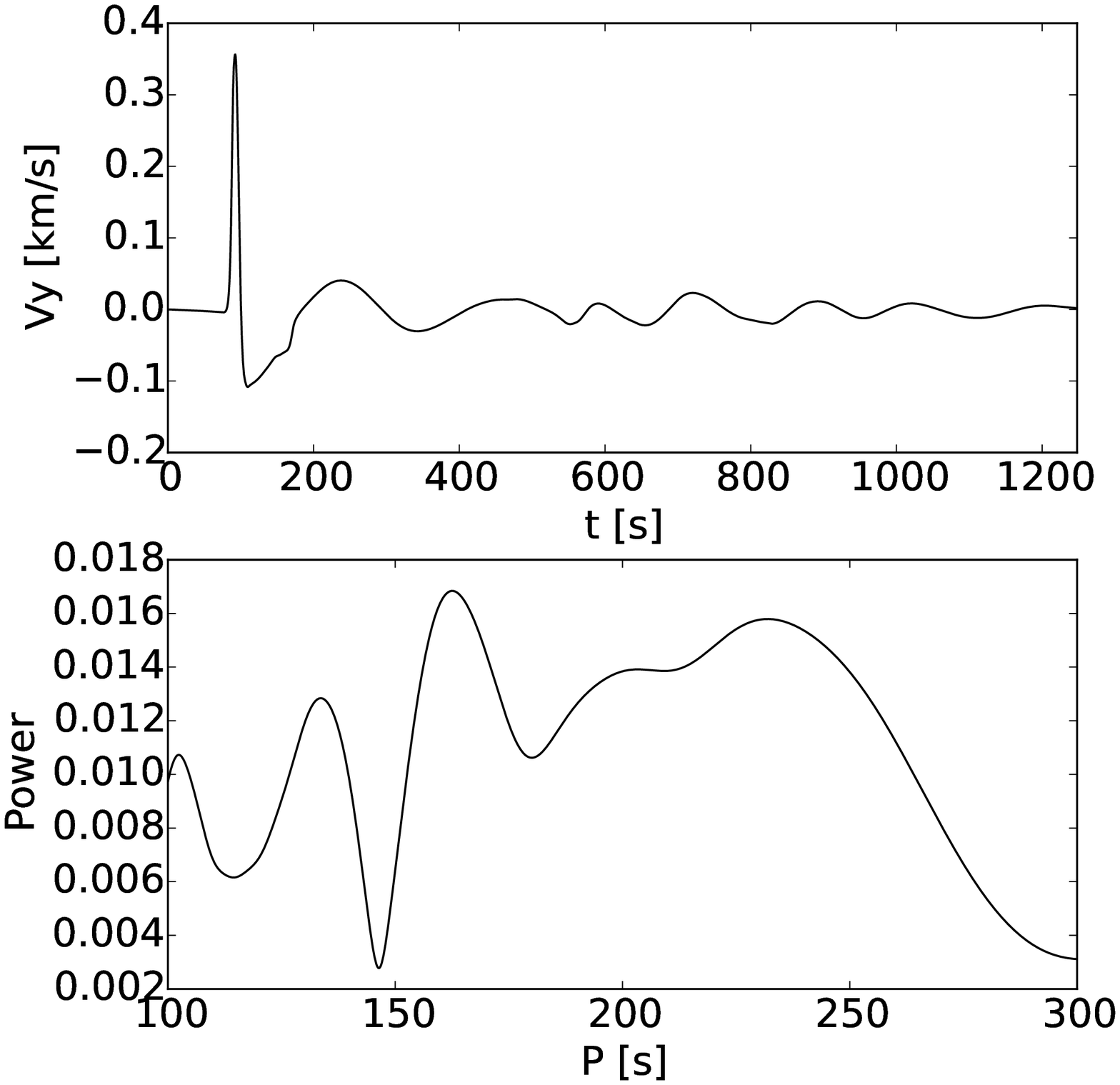}
\vspace*{+2cm}
\caption{\small
Time-signatures (the top panels) and their periodo-grams (the bottom panels) of 
$V_{\rm y}$ collected at $y=0.4$~Mm (the left panel) and $y=0.525$~Mm (the right panel) for 
$A_{\rm v}=0.1$~km s$^{-1}$ and $y_{\rm 0}=-150$ km.
} 
\label{fig:ts}
\end{center}
\end{figure*}

As a result of the initial pulse given by Eq.~(\ref{eq:init_per}), acoustic waves are 
generated (Fig.~\ref{fig:ts}, the top sub-panels) and they propagate in the solar 
atmosphere's model. 

As shown first by Lamb (1909), the presence of gravity leads 
to the appearance of the acoustic cutoff period, $P_{\rm ac}={4 \pi \Lambda}/{c_{\rm s}}$, 
where $\Lambda \sim T$ is the pressure scale height, 
which becomes responsible for the waves to be propagating if their period $P$ is smaller 
than $P_{\rm ac}$, or evanescent if $P$ becomes comparable or larger than $P_{\rm ac}$.
It must be noted that in the realistic solar atmosphere considered here, $P_{\rm ac}$
is a local quantity (e.g., Musielak et al. 2006; Routh \& Musielak 2014) that varies 
significantly with height. Lamb (1909, 1932) also showed that an initial pulse results 
in wavefront which propagates away from the launching region.  The wavefront is followed 
by an oscillating wake, which oscillates at the wave period $P_{\rm ac}$ and whose 
amplitude declines in time. 

We analyze time signal of $V_{\rm y}(y,t)$ that is collected at two altitudes: $y=0.4$~Mm
and $0.525$~Mm (Fig.~\ref{fig:ts}).  The leading wavefront and oscillation wake are clearly 
seen in the time-signatures (the top sub-panels).  These time-signatures are analyzed spectrally 
to obtain power spectra (Fig.~\ref{fig:ts}, the bottom sub-panels) that allow us to determine 
the dominant wave-period $P$ for each detection point.  Note that for $y=0.4$ Mm the maximum 
of $P\approx 220$ s is followed by a smaller local maximum at $P\approx 170$ s. For $y=0.525$ 
Mm; the second local maximum became already the dominant wave period, while the former maximum 
at $P\approx 220$ s is now a local maximum.  This simply means that at $y=0.4$ Mm most of wave 
energy is associated with larger period waves, but right above, mainly at $y=0.525$ Mm the 
shorter wave period waves become dominant. 

\begin{figure}
\begin{center}
\includegraphics[width=10.5cm]{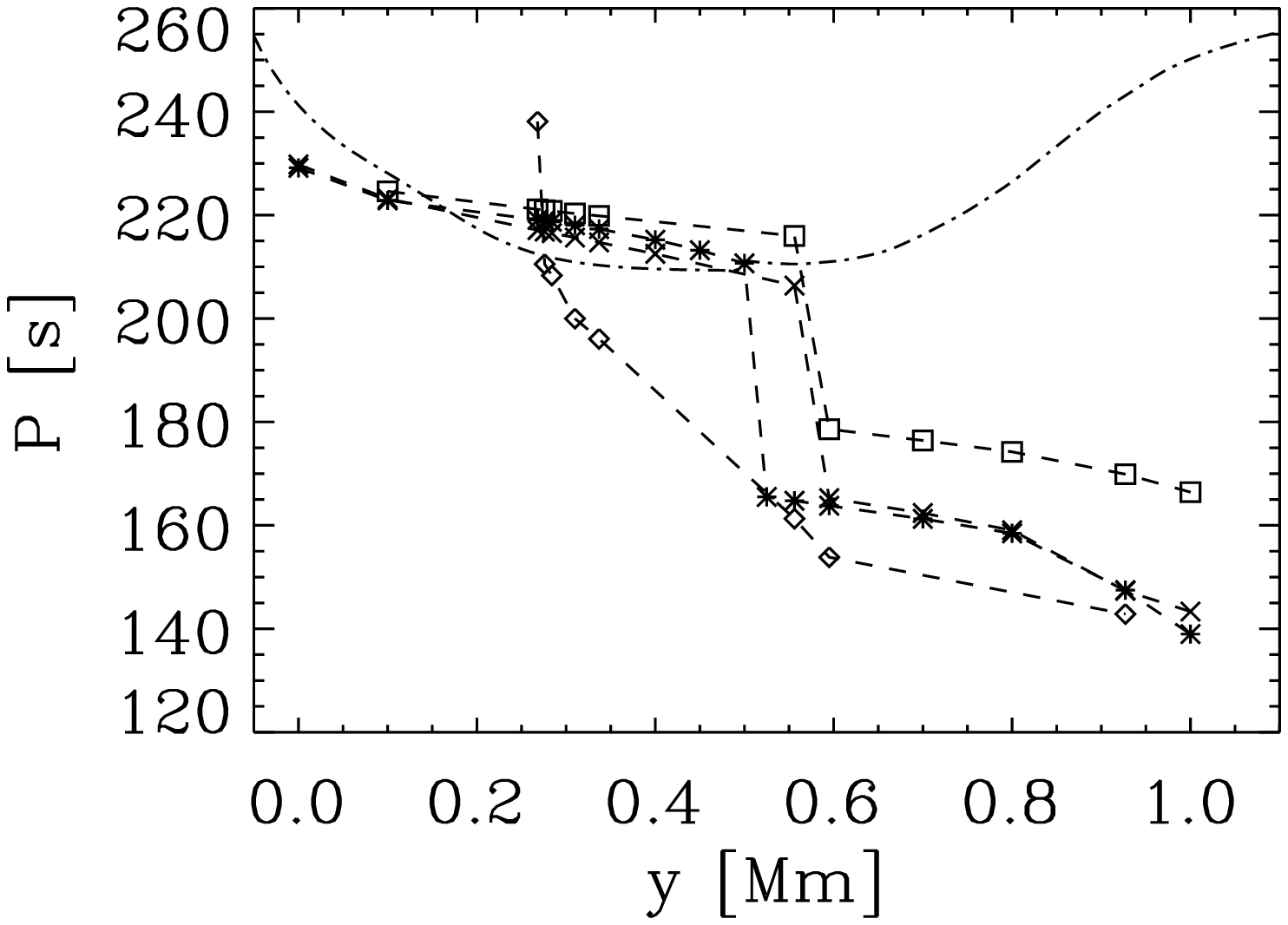}
\caption{\small
The dominant wave-periods vs height $y$ 
for 
$A_{\rm v}=0.1$~km s$^{-1}$  and $y_{\rm 0}=-150$ km (asterisks),
$A_{\rm v}=0.25$~km s$^{-1}$ and $y_{\rm 0}=-150$ km (squares), 
$A_{\rm v}=0.1$~km s$^{-1}$  and $y_{\rm 0}=-250$ km ($\times$),
and 
the observational data of Wi\'sniewska et al. (2016) (diamonds). 
Dashed-dotted line represents acoustic cut-off wave-period, $P_{\rm ac}$. 
} 

\label{fig:Py}
\end{center}
\end{figure}

Figure~\ref{fig:Py} illustrates the numerically evaluated dominant wave period, $P$, which is 
plotted versus altitude $y$, with the observational data of Wi\'sniewska et al. (2016) being
represented by diamonds, and the acoustic cut-off wave period, $P_{\rm ac}$ as dashed-dotted 
line.  The numerical data corresponds to the following cases: 
(a) $A_{\rm v}=0.1$~km s$^{-1}$  and $y_{\rm 0}=-150$ km (asterisks); 
(b) $A_{\rm v}=0.25$~km s$^{-1}$ and $y_{\rm 0}=-150$ km (squares); 
(c) $A_{\rm v}=0.1$~km s$^{-1}$  and $y_{\rm 0}=-250$ km ($\times$). 
Note that the data for the (a) is closest to the observational findings (diamonds). 
Intuitively, we expect that a larger amplitude initial pulse should result in longer wave periods. 
Indeed, Fig.~\ref{fig:Py} confirms that.  For deeper launched pulses, such as in the case of (c), 
larger amplitude oscillating tail is seen, and $P$ corresponding to the case of (c) is larger than 
$P$ for the case of (a).   Moreover, for low values of $y$ the dominant wave periods are shorter 
than $P_{\rm ac}$, and the acoustic waves are propagating in these atmospheric layers. 
However, for $y>0.15$ Mm we find that 
$P>P_{\rm ac}$, and as a result, the acoustic waves become evanescent.  
The long dominant wave period seen in Fig.~1 reduces its magnitude and 
the lagging first local maximum in $P$ becomes dominant for $y=0.525$ Mm; 
this means that much of the energy carried by the waves was converted into short period waves. 
As $P<P_{\rm ac}$ for $y>0.55$ Mm, 
the acoustic waves of such wave periods are propagating in the atmosphere.  
Note that values of $P$ are within the range of about $140$ and $240$ seconds, that corresponds 
to approximately $2.5$- and $4$-minutes oscillations.  The numerically detected wave-periods 
exhibit a fall-off with height for all chosen parameters and the numerical results are close 
to the observational data reported by Wi\'sniewska et al. (2016). 
%
\section{Conclusions}\label{sec:conclusions}
%
In this paper, we simulated numerically the behavior of acoustic waves in low layers 
of the solar atmosphere that is magnetic-free and invariant along horizontal directions. 
Our main goal was to reconcile theory with the most recent observations performed 
by Wi\'sniewska et al. (2016), who demonstrated how the acoustic cutoff varies with 
height in the solar atmosphere.  In our approach, the waves are excited by a single 
initial pulse in vertical component of velocity of its amplitude $V_{\rm y}=0.1$~
km s${}^{-1}$ and $V_{\rm y}=0.25$~km s${}^{-1}$, and the pulse leads to a spectrum 
of acoustic waves of different periods that propagate throughout the background 
solar atmosphere.  During this propagation, the spectrum is filtered by the atmosphere, 
and we used its non-propagating part representing standing acoustic waves to determine 
the resulting acoustic cutoff period, which varies with height. The numerically obtained 
falling-off trend of the dominant wave period generally matches the observational data 
of Wi\'sniewska et al. (2016).  The agreement clearly indicates that the obtained numerical 
results may be used as a basis to determine the structure of the background solar atmosphere. 

Finally, we want to point out that all presented results were obtained with the fixed
$\gamma = 1.4$ and that the OPAL equation of state would lead to the adiabatic index in 
the solar atmosphere varying between $1.1$ and $1.66$ within the first $1.5$ Mm above 
the solar surface.  This variation may affect the acoustic cut-off period by $20$ \% 
or less as compared to the constant gamma case, and will probably lead to the change 
of a similar magnitude in the wave behaviour in our numerical simulations. 

\acknowledgments
The authors are thankful to an anomous referee for his/her comment on 
the earlier versions of this manuscript. 
KM's and ZEM's work was done in the framework of the project from the Polish National 
Foundation (NCN) Grant no. 2014/15/B/ST9/00106.  The work has also been supported by 
NSF under the grant AGS 1246074 (Z.E.M. and K.M.), and by the Alexander von Humboldt 
Foundation (Z.E.M.). 
This work was also supported by the German Academic Exchange Service DAAD (A.W.). 

A.W. acknowledges support from the Solarnet project, which has received funding from 
the European Commission’s FP7 Capacities Programme for 
the period 2013 April - 2017 March under grant agreement No. 312495. 

Numerical simulations were performed on the LUNAR/SOLARIS cluster
at Institute of Mathematics of University of M. Curie-Sk{\l}odowska, Lublin, Poland. 



\clearpage

\clearpage

%
%

%
%


\begin{thebibliography}{}

\bibitem{} Avrett, E.H. \& Loeser, R. 2008, ApJS, 175, 229
\bibitem{} Campos, L.M.B.C., 1986, Rev. Mod. Phys., 58, 117 
\bibitem{} Carlsson, M., Stein, R.F., 1997, ApJ, 481, 500
\bibitem{} Cuntz, M., Ulmscheider, P., Musielak, Z.E., 1998, ApJL, 493, L117
\bibitem{} Fawzy, D.E., Musielak, Z.E., 2012, MNRAS, 421, 159 
\bibitem{} Fawzy, D.E., Rammacher, W., Ulmschneider, P., Musielak, Z.E.,
           St\c epie\'n, K., 2002, A\&A 386, 971
\bibitem{} Fleck, B., Schmitz, F., 1993, A\&A, 273, 487
\bibitem{} Jim\'enez, A., 2006, ApJ, 646, 1398
\bibitem{} Jim\'enez, A., Garc\'ia, R.A., Pall\'e, P.L., 2011, ApJ, 743, 99
\bibitem{} Lamb, H., 1909, Proc. Lond. Math. Soc., 7, 122 
\bibitem{} Lamb, H., 1910, Proc. R. Soc. London, A, 34, 551 
\bibitem{} Lamb, H., 1932, Hydrodynamics (Dover, New York)
\bibitem{} Moore, D.W., Spiegel, E.A., 1964, ApJ., 139, 48 
\bibitem{} Mignone, A., Zanni, C., Tzeferacos, P., van Straalen, B., 
           Colella, P., Bodo, G., 2012, ApJS, 198, 31
\bibitem{} Musielak, Z.E., Musielak, D.E., Mobashi, H., 2006, Phys. Rev. E, 73, 036612-1
\bibitem{} Routh, S., Musielak, Z.E., 2014, Astron. Nachr., 335,1043 
\bibitem{} Souffrin, 1966, AnAp., 39,55 
\bibitem{} Summers, D., 1976, Quart. J. Mech. Appl. Math., 29, 117 
\bibitem{} Ulmschneider, P., 1971, A\&A, 14, 275
\bibitem{} Ulmschneider, P., Schmitz, F., Kalkofen, W., Bohn, H.U., 1978, A\&A, 70, 487
\bibitem{} Wi\'sniewska, A., Musielak, Z.E., Staiger, J., Roth, M., 2016, ApJL, 819, L23 
 
%
\end{thebibliography}
\end{document}